\documentclass[final,5p,twocolumn,times,sort&compress]{elsarticle}

\def\preprintdate{IUHET 612, July 2016}

\usepackage{epsfig}

\def\tfrac{\frac}

\def\be{\beta}

\def\de{\delta}
\def\ep{\epsilon}

\def\ka{\kappa}
\def\la{\lambda}

\def\si{\sigma}
\def\vs{\varsigma}

\def\ph{\phi}

\def\ch{\chi}

\def\om{\omega}

\def\De{\Delta}

\def\La{\Lambda}

\def\Om{\Omega}
\def\mn{{\mu\nu}}

\def\half{{\textstyle{1\over 2}}}
\def\lsim{\mathrel{\rlap{\lower4pt\hbox{\hskip1pt$\sim$}}
    \raise1pt\hbox{$<$}}}
\def\gsim{\mathrel{\rlap{\lower4pt\hbox{\hskip1pt$\sim$}}
    \raise1pt\hbox{$>$}}}
\def\sqr#1#2{{\vcenter{\vbox{\hrule height.#2pt
         \hbox{\vrule width.#2pt height#1pt \kern#1pt
         \vrule width.#2pt}
         \hrule height.#2pt}}}}
\newcommand{\beq}{\begin{equation}}
\newcommand{\eeq}{\end{equation}}
\newcommand{\bea}{\begin{eqnarray}}
\newcommand{\eea}{\end{eqnarray}}
\newcommand{\rf}[1]{(\ref{#1})}
\def\etal{{\it et al.}}

\def\re{{\rm Re}}
\def\im{{\rm Im}}

\def\kem{\hat\ka_{e-}}
\def\kop{\hat\ka_{o+}}

\def\voc{\mathrel{\rlap{\lower0pt\hbox{\hskip1pt{$c$}}}
    \raise3pt\hbox{$\neg$}}}
\def\vok{\mathrel{\rlap{\lower0pt\hbox{\hskip1pt{$k$}}}
    \raise6pt\hbox{$\neg$}}}

\def\cjm#1#2#3{c^{(#1)}_{(#2)#3}}
\def\kI{\cjm{d}{I}{jm}}

\def\kIdjm#1#2{\cjm{#1}{I}{#2}}

\def\MIlab{{\cal M}^{(d)\, \rm lab}_{(I)jm}}

\def\lhat{\hat l}
\def\xhat{\hat l_1}
\def\yhat{\hat l_2}

\def\clabjm#1#2#3{c^{(#1){\rm lab}}_{(#2)#3}}
\def\kIlab{\clabjm{d}{I}{jm}}

\def\nn{\nonumber}

\def\wT{\om_\oplus T_\oplus}
\def\WT{\Om_\oplus T}
\def\C#1{\,\kIdjm4{#1}}
\def\RC#1{\,\re ~\kIdjm4{#1} }
\def\IC#1{\,\im ~\kIdjm4{#1} }
\def\Cs#1{\,\kIdjm6{#1}}
\def\RCs#1{\,\re ~\kIdjm6{#1} }
\def\ICs#1{\,\im ~\kIdjm6{#1} }

\def\kem{\widetilde\ka_{e-}}		
\def\kop{\widetilde\ka_{o+}}		
\def\kt{\widetilde\ka_{\rm tr}}		

\def\ol#1{\overline{#1}{}}
\def\frac#1#2{{\textstyle{{#1}\over {#2}}}}
\def\fr#1#2{{{{#1}\over {#2}}}}

\def\f{f}

\begin{document}

\begin{frontmatter}

\title{Searching for photon-sector Lorentz violation 
using gravitational-wave detectors}

\author{V.\ Alan Kosteleck\'y,$^1$ 
Adrian C.\ Melissinos,$^2$
and Matthew Mewes$^3$}

\address{$^1$Physics Department, Indiana University,
Bloomington, Indiana 47405, USA\\
$^2$Department of Physics and Astronomy, University of Rochester, 
Rochester, New York 14627, USA\\ 
$^3$Physics Department, California Polytechnic State University, 
San Luis Obispo, California 93407, USA}

\address{}
\address{\rm \preprintdate; 
published in Physics Letters B,
DOI:10.1016/j.physletb.2016.08.001}

\begin{abstract}
We study the prospects for using
interferometers in gravitational-wave detectors 
as tools to search for photon-sector violations of Lorentz symmetry.
Existing interferometers are shown to be exquisitely sensitive 
to tiny changes in the effective refractive index of light
occurring at frequencies around and below the microhertz range,
including at the harmonics of the frequencies of the Earth's sidereal rotation 
and annual revolution relevant for tests of Lorentz symmetry.
We use preliminary data obtained 
by the Laser Interferometer Gravitational-Wave Observatory (LIGO) in 2006-2007 
to place constraints on coefficients for Lorentz violation 
in the photon sector exceeding current limits 
by about four orders of magnitude.
\end{abstract}

\end{frontmatter}

Interferometry has been a valuable tool for investigating relativity
for well over a century,
beginning with the classic 
Michelson-Morley and Kennedy-Thorndike experiments
\cite{mm,kt}
that helped to establish the underlying Lorentz symmetry of relativity.
The suggestion that tiny deviations from Lorentz invariance
could arise from an underlying unified theory
such as strings 
\cite{ksp} 
has revitalized experimental efforts to probe relativity in recent years, 
leading to many sensitive searches for Lorentz violation
involving interferometric experiments with light, particles, and atoms
\cite{tables}.
Recently,
the relativistic prediction of gravitational waves
has been confirmed using interferometric techniques 
\cite{ligo-gw}.

The world's largest laser interferometers
are associated with gravitational-wave observatories,
and it is natural to ask about their potential sensitivity
to Lorentz-violating effects involving photons.
Existing observatories include
the Laser Interferometer Gravitational-Wave Observatory (LIGO)
\cite{ligo}
with interferometers located 
at Hanford, Washington and Livingston, Louisiana,
and the Virgo observatory
\cite{virgo}
with interferometer located near Pisa, Italy.
Other large ground-based observatories are operational or planned 
\cite{geo,kagra,ligo-in}, 
and efforts to develop a space-based observatory,
the Laser Interferometer Space Antenna (LISA)
\cite{lisa}
are underway.
Here,
we examine the potential for using low-frequency data 
from these interferometers
to search for signals of Lorentz violation 
in the form of rotation and boost asymmetries 
associated with the sidereal rotation and annual revolution of the Earth. 
We present a general theoretical framework
for discussing the effects,
and we apply it to preliminary LIGO data collected in 2006-2007
with the Hanford instrument.
The results obtained below reveal 
an attained sensitivity to Lorentz violation in the photon sector
about four orders of magnitude
greater than current laboratory experiments.

A rough estimate of the sensitivity of the gravitational-wave instruments 
can be made by noting that each can be idealized 
as a Michelson interferometer with Fabry-P\'erot cavities in the arms.
At LIGO,
for example,
the physical size of each arm is $L \simeq 4$ km,
with an effective path length for the laser light
of about 1000 km due to the cavity finesse $F\simeq 280$
for the configuration during the 2006-2007 run.
The laser operates at an infrared wavelength $\la \simeq 1064$ nm,
and the relative fringe shift $S$ 
can be measured to $S \simeq 4 \times 10^{-10}$.
Taken together, 
these values suggest that an effective sensitivity
to a shift $\de f$ of the frequency $f$ 
of $\de f/f \approx S \la/FL \simeq 4 \times 10^{-22}$
is attainable.
This estimate suggests that gravitational-wave observatories
potentially have intrinsic sensitivities to Lorentz violation 
several orders of magnitude better 
than those achieved in recent Michelson-Morley experiments
\cite{tables,ch16,na15,pr15}.
It thus provides motivation for the present investigation
of the prospects for tests of Lorentz symmetry 
with LIGO and other gravitational-wave interferometers.

The LIGO interferometer is optimized for detection
of gravitational-wave signals in the approximate range 40-1000 Hz.
The measured signal at the detector port
can be taken as the net phase shift 
\beq
\De \ph = \de\ph_1 - \de\ph_2
\eeq
arising from the individual phase shifts $\de\ph_j$, $j=1,2$,
experienced by the light in each of the two arms.
These individual phase shifts can in principle arise from
changes $\de L_j$ in the effective path lengths $L_j$ of the arms,
or a change $\de \f_c$ in the carrier frequency $\f_c$ of the light.
The phase shifts can also be affected by modifications $\de \ol n_j$
of the effective refractive index $\ol n$ 
experienced by the light propagating in the two arms,
including changes that might arise due to the presence of Lorentz violation.
The net phase shift on the $j$th arm 
for a single light traversal of length $2L$ 
can thus be expressed as
\beq
\fr {\de \ph_j}{2\pi} = \left(
\fr {\de L_j} {L} 
+\fr {\de \f_c} {\f_c} 
+\fr {\de \ol n_j} {\ol n} 
\right)
\fr{2L}{\la}.
\eeq

In its operating mode,
the interferometer is `locked' on a dark fringe 
by adjusting the carrier frequency $\f_c$
and the effective path lengths $L_j$
using feedback and servo mechanisms, 
so that $\De \ph = 0$ is enforced at the detector port 
in the absence of a gravitational-wave signal.
Over a sufficiently large time interval $T$
compared to the time between successive feedback and servo actions,
the integrated net phase change 
reduces to an integral over changes 
in the difference $\De \ol n = \de \ol n_1 - \de \ol n_2$
of effective refractive indices,
\beq
\int_{t-T/2}^{t+T/2} dt  \fr {\De \ph_j}{2\pi} 
\to 
\fr{2L}{\la}
\int_{t-T/2}^{t+T/2} dt  
\fr {\De \ol n} {\ol n} ,
\label{harmonic}
\eeq
because the changes $\de L_j$ and $\de \f_c$ are stochastic
and average to zero when the interferometer is locked.

The above reasoning demonstrates that 
the operating mode of the interferometer 
does in principle have sensitivity 
to time-varying signals from Lorentz violation 
in the effective refractive index $\ol n$.
However,
the Earth's sidereal-rotation angular frequency is 
$\om_\oplus 
\simeq 7.3 \times 10^{-5}$ rad Hz,
while its annual-revolution angular frequency is 
$\Om_\oplus 
\simeq 2.0 \times 10^{-7}$ rad Hz,
so the sidereal and annual signals 
of interest for searches for Lorentz violation
involve frequencies many orders of magnitude
below the optimized band of the instrument.
At these low frequencies,
the instrumental noise makes clean extraction of any signal challenging.
One possible option for sidestepping this issue
is to take advantage of information circulating in the interferometer
at sideband frequencies,
as we discuss next.

The arms are in resonance when the carrier frequency $\f_c$
takes the value $\f_c = N \f_{\rm fsr}$,
where $N$ is typically a large integer of order $10^{10}$
and $\f_{\rm fsr} = c/2L \simeq 37.52$ kHz at LIGO
is called the free spectral range (fsr) frequency.
Resonance also occurs 
at the sidebands $\f_\pm = \f_c \pm \f_{\rm fsr}$,
which experience lower noise
and are thus interesting candidates for signal analysis.
Furthermore,
a {\it macroscopic} difference $\De L = L_1 - L_2$
in the arm lengths,
which for LIGO is of order 2 cm,
displaces these sidebands from the dark fringe
by a bias phase shift
$\ph_b = \pm {\De L}/{2L}
\simeq 3\times 10^{-6}$ 
per traversal of the light.
This implies that the power at the detector port
at the sideband frequency $\f_+$
contains an interference term
between the bias phase shift
and any phase shift from the change \rf{harmonic}
in the difference $\De\ol n$ of effective refractive indices.
The power at $\f_{+}$ is thus modulated 
at the frequencies of harmonic changes in $\De\ol n$.
In short,
when the carrier frequency is used to lock the interferometer,
the sideband at the fsr frequency
can be used to measure the low-frequency signals 
from Lorentz violation
\cite{am10,gu14}. 

A successful measurement of harmonic changes in $\De \ol n$
associated with tidal forces has already been demonstrated
\cite{am14}.
The tidal acceleration has a gravity-gradient component $g_h$ 
along the interferometer arms
that induces redshifts in the circulating light.
The redshifts act to produce effective changes in $\De \ol n$
varying harmonically at the tidal frequencies
and introduce a single-traversal phase shift of
\beq
\fr{\de \ph}{2 \pi} = \fr {g_h L^2}{\la c^2} . 
\eeq
Using preliminary LIGO data from the 2006-2007 run
\cite{am10},
the spectral powers at the tidal frequencies 
are found to be in approximate agreement with results
from standard modeling of the tidal gradients.
Note that the tidal frequencies also appear 
in the demodulated carrier signal 
at the detector port but are compromised by noise,
while they are observable at their exact frequencies
in the spectrum obtained from the fsr signal.

To investigate possible signals from Lorentz violation,
we adopt here the methods of effective field theory,
which provide powerful and model-independent techniques
for studying observable signals originating 
from an otherwise unattainable large energy scale
\cite{sw}. 
The realistic effective field theory describing general Lorentz violation
is called the Standard-Model Extension (SME)
\cite{ck,akgrav}.
It is constructed by adding Lorentz-violating terms
to the action for General Relativity coupled to the Standard Model.
Each addition to the Lagrange density
is a coordinate-independent contraction
of a Lorentz-violating operator with a coefficient 
determining the size of its physical effects.
Any operator can be classified according to its mass dimension $d$
in natural units,
with the corresponding coefficient having mass dimension $4-d$.
Operators of larger $d$ can plausibly be interpreted
as representing effects at higher order
in a low-energy expansion of the underlying theory.
In Minkowski spacetime,
limiting attention to terms with $d\leq 4$
produces a theory that is power-counting renormalizable
and known as the minimal SME.
Reviews include,
for example,
Refs.\ \cite{tables,rb,jt}.

In the present work,
we focus attention on possible effects from the photon sector
of the SME.
We analyze potential signals at harmonics
of the sidereal frequency $\om_\oplus$
and the annual frequency $\Om_\oplus$,
including the sidebands.
In principle,
Lorentz-violating contributions to the signal
could also arise from the matter sector,
including in particular from the electrons, protons, and neutrons
in the interferometer mirrors.
While of definite interest,
addressing this possibility would complicate the present analysis 
without contributing to our goal
of demonstrating that gravitational-wave detectors 
have competitive sensitivity to Lorentz violation,
and so we defer it to future investigation.
This obviates the issue of fixing possible field redefinitions
and coordinate choices
\cite{ck,akgrav,redefs,km09,km02}. 
We also simplify the analysis by disregarding
contributions to Lorentz-violating birefringence of light,
as disentangling these effects
requires unavailable information about the polarization
of the light circulating in the interferometer.

The possible modifications to the effective refractive index
for photons propagating in the presence of Lorentz violation
have been classified and enumerated for arbitrary $d$
\cite{km09}.
Nonbirefringent Lorentz-violating operators in the photon sector
appear only for even $d\geq 4$.
Decomposing in spherical harmonics implies 
the corresponding spherical coefficients for Lorentz violation
can be denoted by $\kI$,
where the subscript $I$ indicates nonbirefringence
and the indices $jm$ are the usual angular quantum numbers 
for the spherical harmonics with $j\leq d-2$.
All the associated modifications to the effective refractive index
can then be expressed in the form
\cite{km09}
\beq
n = 1 + \vs^0,
\quad
\vs^0 = \sum_{djm} E^{d-4} (-1)^j\,
Y_{jm}(\hat l)\, \kIlab,
\eeq
where $E$ is the photon energy,
$\hat l$ is the direction of its momentum,
$\kIlab$ are the coefficients for Lorentz violation
seen in the laboratory frame,
and $d\geq 4$ takes only even values.

To apply the above results in the context of LIGO,
consider first a single arm of the interferometer.
For a traversal of the light down the arm and back,
we can introduce an averaged refractive index
\bea
\ol n(\lhat) &=& \half\big(\vs^0(\lhat)+\vs^0(-\lhat)\big)
\nn\\
&=&
1 + \sum_{djm} E^{d-4}\,
\half\big(1+(-1)^j\big)\, 
Y_{jm}(\lhat)\, \kIlab.
\eea
Taking both arms into account,
the difference $\De\ol n$ appearing in Eq.\ \rf{harmonic}
is then given by
\beq
\De\ol n = \ol n(\xhat) - \ol n(\yhat),
\label{den}
\eeq
where the angle between $\xhat$ and $\yhat$ can be taken as $\pi/2$. 

The LIGO observatory is a noninertial frame
due to the rotation and revolution of the Earth.
In searching for Lorentz violation,
it is useful to work instead in a frame
that is approximately inertial over the time period of the experiment.
The canonical choice for this inertial frame
is the Sun-centered frame
\cite{tables,km02,sunframe},
with coordinates denoted as $(T,X,Y,Z)$.
The origin of the time $T$ is defined 
to be the vernal equinox 2000,
so that $T$(2000-03-20 07:35 UTC)$=0$.
The $Z$ axis is aligned with the Earth's rotation axis,
and the $X$ axis points towards the vernal equinox 2000.
The coefficients $\kI$ can plausibly be assumed constant
on solar-system scales in this frame
\cite{ck}.
The rotation and revolution of the Earth
thus induce sidereal and annual variations
in the laboratory coefficients $\kIlab$.
These variations are key signals for detecting Lorentz violation.

Consider first sidereal variations.
The spherical coefficients $\kI$ for Lorentz violation
are particularly well suited for studies of sidereal signals
because they transform under rotations in a comparatively simple way.
The relationship between the spherical coefficients
in the laboratory frame and ones in the Sun-centered frame
is given by
\cite{km09} 
\beq
\kIlab
=\sum_{m'} e^{im'\om_\oplus T_\oplus} d^j_{mm'}(-\ch) \kIdjm d {j m'},
\label{rottr}
\eeq
where $\ch$ is the colatitude of the laboratory
and the little Wigner matrices $d^j_{mm'}$
are specified in Eq.\ (136) of Ref.\ \cite{km09}.
The time $T_\oplus = T-T_0$ is a local sidereal time,
offset from $T$ by
$T_0 \simeq (23.934~{\rm hr})(66.25^\circ- \la)/360^\circ$,
where $\la$ is the longitude of the laboratory in degrees.
For the Hanford site,
$\ch \simeq 43.5^\circ$ 
and $T_0\simeq$ 2000-03-20 19:56 UTC.

Substituting the result \rf{rottr} into the difference \rf{den} gives 
\beq
\De\ol n =
\sum_{djmm'} \MIlab\,
e^{im'\om_\oplus T_\oplus} d^j_{mm'}(-\ch)\,
\kIdjm{d}{jm'}.
\label{nsidereal}
\eeq
In this expression,
the experiment-dependent factor $\MIlab$ is given by
\beq
\MIlab =  E^{d-4}\,
\half\big(1+(-1)^j\big)\,(1- i^m)\,
Y_{jm}(\tfrac{\pi}{2},\ph),
\eeq
where $\ph$ is the angle of the interferometer `$X$' arm
measured east of south,
which is $\ph \simeq -144^\circ$ for the interferometer 
at the Hanford site.
As an example,
Table 1 displays the explicit numerical form 
of the combinations \rf{nsidereal} for harmonics with $d=4$ and $d=6$
for this site.
The first column shows the harmonic.
The second column contains the combination for $d=4$
contributing to the difference $\De\ol n$.
The third column lists the combinations contributing for $d=6$. 
The numerical factors in this last column
are given in units of $10^{-18}$ GeV$^2$.
The contributions to $\De\ol n$ from an individual harmonic
can be obtained from this table by multiplying
an entry in the first column with one in the second or third column.

Next,
consider annual variations.
These are associated with boosts
between the Sun-centered and laboratory frames,
so working with cartesian coefficients for Lorentz violation
is conceptually more straightforward than spherical coefficients.
To keep the analysis comparatively simple
we focus here on the case $d=4$,
for which the effects are unsuppressed by powers of the energy $E$.
A more general analysis is possible in principle
and would be of interest
but lies beyond our present scope.

In cartesian coordinates and for $d=4$,
the modification to the effective refractive index in the laboratory frame
can be written as
\beq
\vs^0 = -\half \lhat^j\lhat^k \kem^{jk} 
+ \half \ep^{jkl}\lhat^j \kop^{kl} 
+ \kt^{\rm lab} ,
\eeq
where the ten cartesian coefficients for Lorentz violation
associated with nonbirefringent operators at $d=4$,
which are linearly related to the spherical coefficients,
are taken as the symmetric combination $\kem^{JK}$, 
the antisymmetric combination $\kop^{JK}$, 
and the trace component $\kt$
in the Sun-centered frame
\cite{km02}.
This gives
\bea
\De\ol n
&=& 
-\half (\lhat_1^j\lhat_1^k - \lhat_2^j\lhat_2^k) \kem^{jk}
\nn\\
&=&
-\half (\lhat_1^j\lhat_1^k -\lhat_2^j\lhat_2^k)
\big(
{\La^j}_J {\La^k}_K \kem^{JK}
+ {\La^j}_T {\La^k}_J \ep^{JKL} \kop^{KL}
\nn\\
&&
\hskip 50pt
-2{\La^j}_T {\La^k}_T \kt
\big).
\eea
In this expression,
the elements 
of the Lorentz transformation
relating the Sun-centered frame and the laboratory frame 
can be taken as
\beq
\hskip -12pt
{\La^0}_T = 1 , 
\quad
{\La^0}_J = -\be^J ,
\quad
{\La^j}_T = -(R\cdot\vec\be)^j ,
\quad
{\La^j}_J = R^{jJ},
\eeq
where the matrix $R^{jJ}$ rotating between the Sun-centered 
and laboratory frames
is given by Eq.\ (C1) of Ref.\ \cite{km02},
and $\be^J$ is given in terms
of the orbital and laboratory boosts
by Eq.\ (C2) of the same reference.

The above set of equations suffices to determine
the explicit form of $\De\ol n$
in terms of $d=4$ cartesian coefficients for Lorentz violation,
once the location and relevant properties of the observatory are specified.
The cartesian coefficients can then be transformed
into spherical ones if desired. 
For example,
for the Hanford site
the explicit contributions for each harmonic
in terms of spherical coefficients for Lorentz violation
in the Sun-centered frame
are displayed in Table 2.
In this table,
the first column specifies the harmonic.
All relevant harmonics involving the sidereal and annual frequencies,
including their sidebands, 
are considered.
The second column gives the parity-even contributions,
which match those shown in Table 1.
The parity-odd contributions,
which are proportional to one power of the boost,
are presented in the third column.
The final column
provides the contributions involving the isotropic coefficient $\C{00}$,
all of which are parity even and involve two powers of the boost.
Note that all nine independent components $\C{\mn}$ appear.
However,
the component $\C{20}$ contributes only to the constant term,
which lacks a characteristic time variation
and can therefore be expected to be more challenging to detect.
Note also that the sole contribution to the twice-annual harmonic
comes from the isotropic coefficient.

\begin{table*}
\begin{center}
\setlength{\tabcolsep}{5pt}
\renewcommand{\arraystretch}{1.4}
\begin{tabular}{c|c|c}
	Harmonic		&	$	d=4  {\rm ~contributions}	$	&	$	d=6 {\rm ~contributions}~ ( \times 10^{-18}$ GeV$^2)	$	\\	\hline\hline
$	1	$	&	$	0.14\C{20}	$	&	$	0.19\Cs{20}- 0.28\Cs{40}	$	\\	
$	\cos(\wT)	$	&	$	0.24\RC{21}-1.0\IC{21}	$	&	$	0.32\RCs{21}+0.062\RCs{41}-1.4\ICs{21}+1.1\ICs{41}	$	\\	
$	\sin(\wT)	$	&	$	-1.0\RC{21}-0.24\IC{21}	$	&	$	-1.4\RCs{21}+1.1\RCs{41}-0.32\ICs{21}-0.062\ICs{41}	$	\\	
$	\cos(2\wT)	$	&	$	0.36\RC{22}-1.1\IC{22}	$	&	$	0.49\RCs{22}+0.061\RCs{42}-1.4\ICs{22}-0.82\ICs{42}	$	\\	
$	\sin(2\wT)	$	&	$	-1.1\RC{22}-0.36\IC{22}	$	&	$	-1.4\RCs{22}-0.82\RCs{42}-0.49\ICs{22}-0.061\ICs{42}	$	\\	
$	\cos(3\wT)	$	&	$	$--$	$	&	$	0.27\RCs{43}-0.64\ICs{43}	$	\\	
$	\sin(3\wT)	$	&	$	$--$	$	&	$	-0.64\RCs{43}-0.27\ICs{43}	$	\\	
$	\cos(4\wT)	$	&	$	$--$	$	&	$	-0.27\RCs{44}+0.78\ICs{44}	$	\\	
$	\sin(4\wT)	$	&	$	$--$	$	&	$	0.78\RCs{44}+0.27\ICs{44}	$	\\	
\end{tabular}
\caption{Contributions from sidereal harmonics 
for $d=4$ and $d=6$ at Hanford, WA.}
\end{center}
\end{table*}

\begin{table*}
\begin{center}
\setlength{\tabcolsep}{5pt}
\renewcommand{\arraystretch}{1.2}
\begin{tabular}{c|c|c|c}
	Harmonic		&		Parity-even		&		Parity-odd ($\times 10^{-5}$)		&		Isotropic ($\times 10^{-10}$)		\\	\hline\hline
$	1	$	&	$	0.14\C{20}	$	&	$	0.068\C{10}	$	&	$	-1.2\C{00}	$	\\	
$	\cos(\WT)	$	&	$	$--$	$	&	$	0.56\C{10}-0.92\IC{11}	$	&	$	0.12\C{00}	$	\\	
$	\sin(\WT)	$	&	$	$--$	$	&	$	-1.0\RC{11}	$	&	$	$--$	$	\\	
$	\cos(2\WT)	$	&	$	$--$	$	&	$	$--$	$	&	$	0.36\C{00}	$	\\	
$	\sin(2\WT)	$	&	$	$--$	$	&	$	$--$	$	&	$	$--$	$	\\	\hline
$	\cos(\wT-2\WT)	$	&	$	$--$	$	&	$	$--$	$	&	$	-5.2\C{00}	$	\\	
$	\sin(\wT-2\WT)	$	&	$	$--$	$	&	$	$--$	$	&	$	-1.2\C{00}	$	\\	
$	\cos(\wT-\WT)	$	&	$	$--$	$	&	$	-6.1\C{10}+0.42\RC{11}-1.8\IC{11}	$	&	$	0.13\C{00}	$	\\	
$	\sin(\wT-\WT)	$	&	$	$--$	$	&	$	-1.4\C{10}-1.8\RC{11}-0.42\IC{11}	$	&	$	-0.30\C{00}	$	\\	
$	\cos(\wT)	$	&	$	0.24\RC{21}-1.0\IC{21}	$	&	$	0.10\RC{11}+0.046\IC{11}	$	&	$	-5.0\C{00}	$	\\	
$	\sin(\wT)	$	&	$	-1.0\RC{21}-0.24\IC{21}	$	&	$	0.046\RC{11}-0.10\IC{11}	$	&	$	-1.2\C{00}	$	\\	
$	\cos(\wT+\WT)	$	&	$	$--$	$	&	$	0.26\C{10}+0.42\RC{11}-1.8\IC{11}	$	&	$	-0.0057\C{00}	$	\\	
$	\sin(\wT+\WT)	$	&	$	$--$	$	&	$	0.062\C{10}-1.8\RC{11}-0.42\IC{11}	$	&	$	0.013\C{00}	$	\\	
$	\cos(\wT+2\WT)	$	&	$	$--$	$	&	$	$--$	$	&	$	0.22\C{00}	$	\\	
$	\sin(\wT+2\WT)	$	&	$	$--$	$	&	$	$--$	$	&	$	0.053\C{00}	$	\\	\hline
$	\cos(2\wT-2\WT)	$	&	$	$--$	$	&	$	$--$	$	&	$	-4.5\C{00}	$	\\	
$	\sin(2\wT-2\WT)	$	&	$	$--$	$	&	$	$--$	$	&	$	13\C{00}	$	\\	
$	\cos(2\wT-\WT)	$	&	$	$--$	$	&	$	-9.1\RC{11}-3.1\IC{11}	$	&	$	$--$	$	\\	
$	\sin(2\wT-\WT)	$	&	$	$--$	$	&	$	-3.1\RC{11}+9.1\IC{11}	$	&	$	$--$	$	\\	
$	\cos(2\wT)	$	&	$	0.36\RC{22}-1.1\IC{22}	$	&	$	$--$	$	&	$	0.39\C{00}	$	\\	
$	\sin(2\wT)	$	&	$	-1.1\RC{22}-0.36\IC{22}	$	&	$	$--$	$	&	$	-1.1\C{00}	$	\\	
$	\cos(2\wT+\WT)	$	&	$	$--$	$	&	$	0.39\RC{11}+0.13\IC{11}	$	&	$	$--$	$	\\	
$	\sin(2\wT+\WT)	$	&	$	$--$	$	&	$	0.13\RC{11}-0.39\IC{11}	$	&	$	$--$	$	\\	
$	\cos(2\wT+2\WT)	$	&	$	$--$	$	&	$	$--$	$	&	$	-0.0083\C{00}	$	\\	
$	\sin(2\wT+2\WT)	$	&	$	$--$	$	&	$	$--$	$	&	$	0.024\C{00}	$	\\	
\end{tabular}
\caption{Contributions from $d=4$ spherical coefficients at Hanford, WA.}
\end{center}
\end{table*}

\begin{figure}
\includegraphics[width=\hsize]{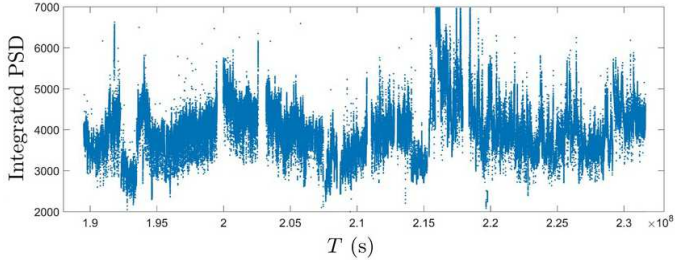}
\caption{
Integrated fsr PSD as a function of time $T$ in the Sun-centered frame.
\label{data}}
\end{figure}

To investigate the experimental reach attainable in practice,
we analyze the preliminary dataset 
taken in the fsr channel
at the Hanford site during the S5 LIGO run,
over the 16-month period from March 31, 2006 to July 31, 2007
\cite{am10}.
During this run,
the photodetector signal was demodulated at 37.52 kHz.
The power spectral density (PSD),
which is proportional to 
the absolute value of the electric-field amplitude squared per Hz,
was evaluated over intervals of 64 s
and then integrated in the range $37.52\pm 0.2$ kHz,
thereby yielding a time series of the power at the fsr frequency.
Figure \ref{data} shows this series.
The vertical axis is the uncalibrated integrated PSD,
while the horizontal axis is the time $T$
in seconds since the vernal equinox 2000.
The time series corresponds to the squared modulus
$| \ph_b + \ph_s|^2$,
where $\ph_b$ is the bias phase shift mentioned above
and $\ph_s$ is the time-dependent phase shift
induced by the time variations in $\De\ol n$.
Note that these data provide an essentially continuous record
over the 16-month period.
This represents another advantage of the fsr channel
in that it provides continuity over this extended period,
whereas the carrier channel is reset
after the interferometer loses lock,
typically after about 24 hours. 

\begin{figure}
\includegraphics[width=\hsize]{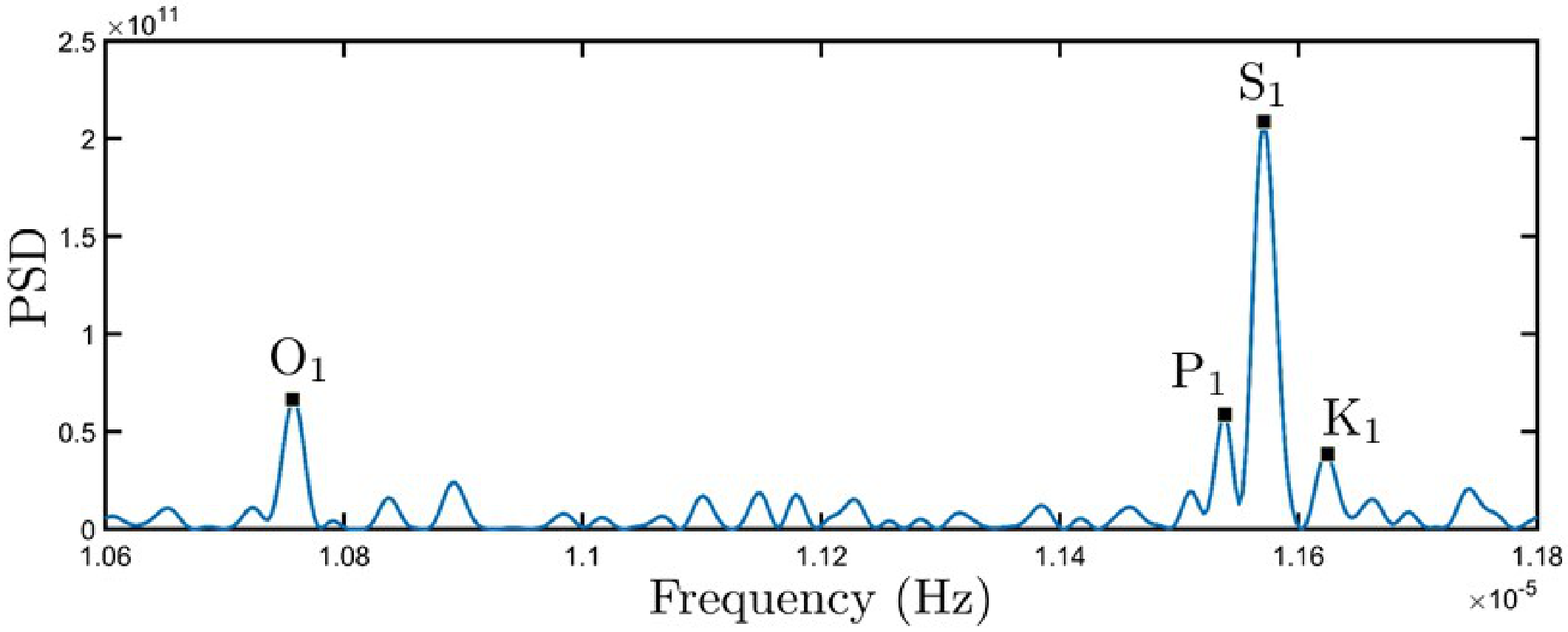}
\caption{
PSD versus frequency in the sidereal region.
\label{once}}
\vskip 20pt
\includegraphics[width=\hsize]{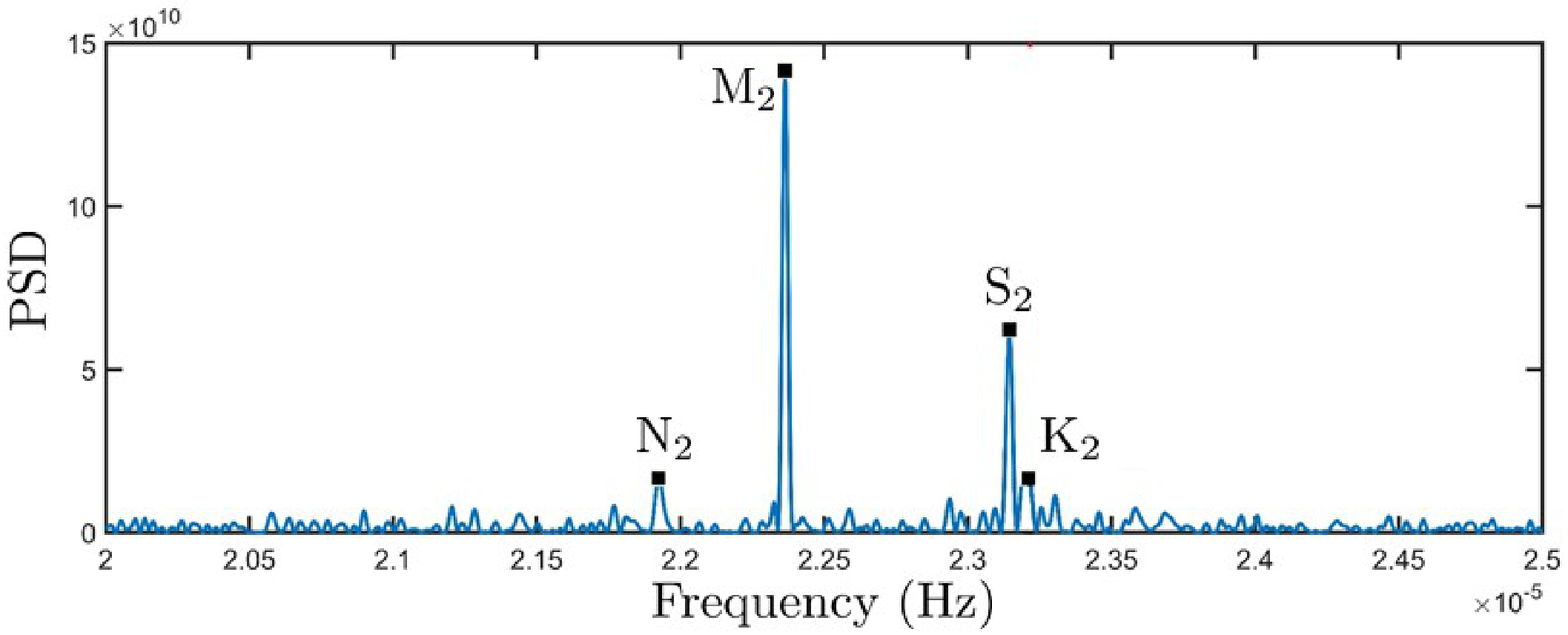}
\caption{
PSD versus frequency in the semisidereal region.
\label{twice}}
\vskip 20pt
\includegraphics[width=\hsize]{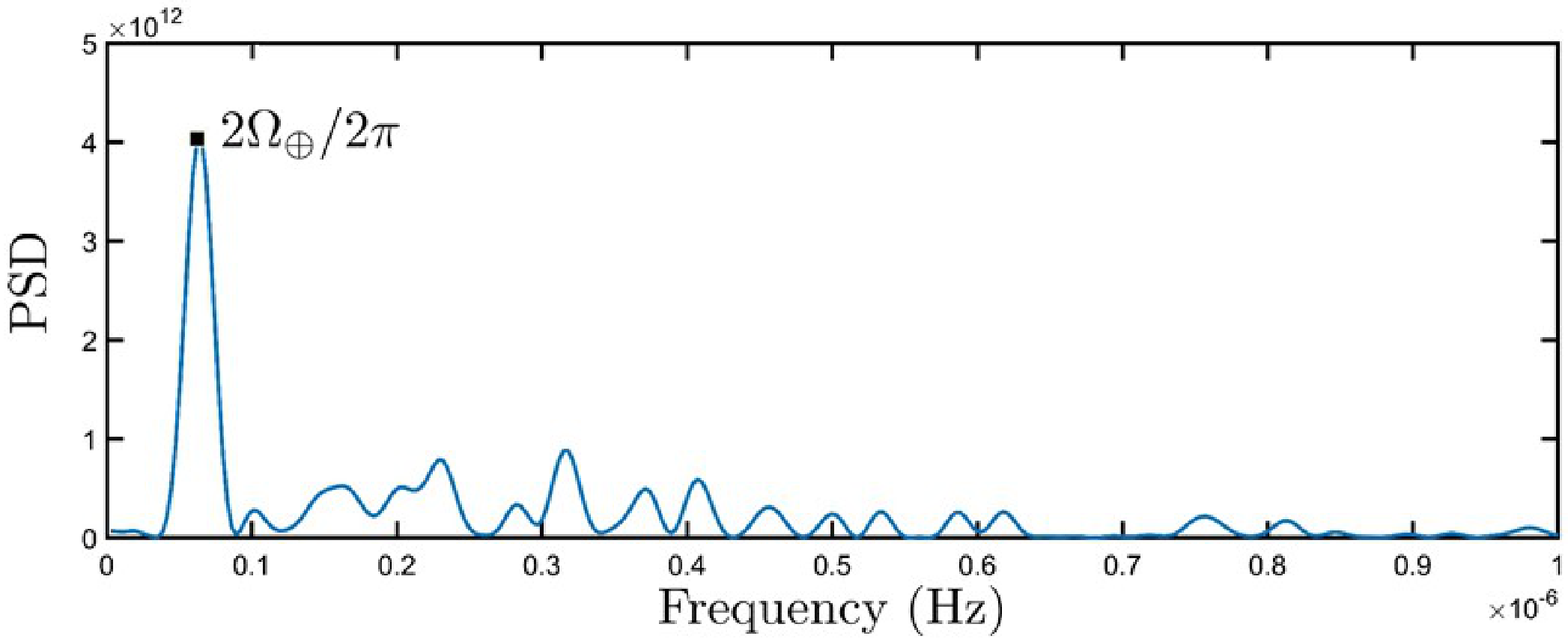}
\caption{
PSD versus frequency in the annual region.
\label{solar}}
\end{figure}

To study the various sidereal and annual signals,
the power spectra in the appropriate frequency ranges
can be extracted from the dataset.
The resolution bandwidth is approximately $2.4\times 10^{-8}$ Hz.
The PSD as a function of frequency in the sidereal region
is shown in Fig.\ \ref{once}.
Table 3 lists the frequencies and the PSD values 
for each of the four prominent peaks.
Four tidal lines are known to appear in this region: 
the lunar principle wave O$_1$,
the solar principle wave P$_1$,
the lunar and solar declinational waves K$_1$,
and the solar elliptical wave S$_1$ of K$_1$.
Near the twice-sidereal frequency,
the power spectrum is presented in Fig.\ \ref{twice},
and the locations and sizes of the four prominent peaks
are provided in Table 3.
Again,
four tidal lines are known here:
the lunar principle wave M$_2$,
the solar principle wave S$_2$,
the lunar major elliptical wave N$_2$ of M$_2$,
and the lunar and solar declinational waves K$_2$.
With one exception,
the frequencies of the four prominent peaks in each of these spectra
match the locations of these tidal lines to $10^{-8}$ Hz.
The measured power in each line is proportional to 
the tidal amplitude because it arises from interference, 
and the observed relative amplitudes agree with known values
\cite{am14}.
The exception is the S$_1$ line,
which is shifted by about 2.5 standard deviations
from the expected frequency
and should be unobservable. 
This line must therefore be attributed to human activities on a daily cycle.

\begin{table}
\begin{center}
\setlength{\tabcolsep}{5pt}
\renewcommand{\arraystretch}{1.2}
\begin{tabular}{c|c|c}
	Peak		&		Frequency		&		Power spectral density		\\	\hline\hline	
$	O_1	$	&	$	1.076\times 10^{-5}	$	&	$	6.655\times 10^{10}	$	\\		
$	P_1	$	&	$	1.154\times 10^{-5}	$	&	$	5.869\times 10^{10}	$	\\		
$	S_1	$	&	$	1.157\times 10^{-5}	$	&	$	2.088\times 10^{11}	$	\\		
$	K_1	$	&	$	1.162\times 10^{-5}	$	&	$	3.841\times 10^{10}	$	\\		
$	N_2	$	&	$	2.192\times 10^{-5}	$	&	$	1.666\times 10^{10}	$	\\		
$	M_2	$	&	$	2.236\times 10^{-5}	$	&	$	1.415\times 10^{11}	$	\\		
$	S_2	$	&	$	2.315\times 10^{-5}	$	&	$	6.218\times 10^{10}	$	\\		
$	K_2	$	&	$	2.321\times 10^{-5}	$	&	$	1.657\times 10^{10}	$	\\		
$	2\Om_\oplus/2\pi	$	&	$	6.239\times 10^{-8}	$	&	$	4.034\times 10^{12}	$	\\		
\end{tabular}
\caption{Frequencies (Hz) and power spectral density of selected peaks.}
\end{center}
\end{table}

The PSD in the vicinity of the solar frequency 
is displayed in Fig.\ \ref{solar}.
No significant annual modulation appears in the data.
However,
a pronounced peak is visible at
the frequency $\f = (6.5\pm 0.6)\times 10^{-8}$ Hz,
which is consistent with the semiannual frequency 
$2\Om_\oplus/2\pi\simeq 6.2 \times 10^{-8}$ Hz.
The amplitude of the declinational solar tidal wave at this frequency
is too small by more than an order of magnitude
to account for this peak,
which has height as shown in Table 3.
The origin of this anomalous peak is currently unknown 
but could be instrumental.
As an illustration of principle,
consider the feed-forward servo mechanism 
that helps to maintain the interferometer lock
by correcting for the tidal deformation of the Earth 
via actuators that modify the macroscopic arm-length difference.
This servo includes a correction at twice the annual frequency,
which conceivably could be a natural source of an instrumental effect.
However, 
in practice the tidal servo would have been reset 
between the lock periods roughly once a day, 
and moreover the size of the correction is too small 
by more than an order of magnitude, 
so this appears unlikely to be 
the source of the observed continuous modulation.
In the analysis that follows,
we include the anomalous peak for completeness,
but its definitive interpretation and verification 
must await the acquisition of an independent dataset.

\begin{table}
\begin{center}
\setlength{\tabcolsep}{5pt}
\renewcommand{\arraystretch}{1.2}
\begin{tabular}{c|c}
	Harmonic		&	$	\De\ol n/ \ol n	$	\\	\hline\hline
$	\om_\oplus	$	&	$	<1.4\times 10^{-20}	$	\\	
$	2\om_\oplus	$	&	$	<2.0\times 10^{-22}	$	\\	
$	3\om_\oplus	$	&	$	<2.1\times 10^{-22}	$	\\	
$	4\om_\oplus	$	&	$	<2.1\times 10^{-22}	$	\\	
$	\Om_\oplus	$	&	$	<3.4\times 10^{-20}	$	\\	
$	2\Om_\oplus	$	&	$	(4.0\pm 0.25)\times 10^{-19}	$	\\	
\end{tabular}
\caption{Results for $\De\ol n /n$ from Hanford, WA preliminary data.}
\end{center}
\end{table}

\begin{table}
\begin{center}
\setlength{\tabcolsep}{5pt}
\renewcommand{\arraystretch}{1.2}
\begin{tabular}{c|c|c}
	Harmonic		&		Coefficient		&		Result			\\	\hline\hline
$	\om_\oplus	$	&	$	|\C{21}|	$	&	$	<1.3\times 10^{-20}	$		\\	
$		$	&	$	|\Cs{21}|	$	&	$	<1.0\times 10^{-2}	$	GeV$^{-2}$	\\	
$		$	&	$	|\Cs{41}|	$	&	$	<1.3\times 10^{-2}	$	GeV$^{-2}$	\\	[5pt]
$	2\om_\oplus	$	&	$	|\C{22}|	$	&	$	<1.8\times 10^{-22}	$		\\	
$		$	&	$	|\Cs{22}|	$	&	$	<1.3\times 10^{-4}	$	GeV$^{-2}$	\\	
$		$	&	$	|\Cs{42}|	$	&	$	<2.4\times 10^{-4}	$	GeV$^{-2}$	\\	[5pt]
$	3\om_\oplus	$	&	$	|\Cs{43}|	$	&	$	<3.0\times 10^{-4}	$	GeV$^{-2}$	\\	[5pt]
$	4\om_\oplus	$	&	$	|\Cs{44}|	$	&	$	<2.6\times 10^{-4}	$	GeV$^{-2}$	\\	[5pt]
$	\Om_\oplus	$	&	$	|\C{00}|	$	&	$	<3.3\times 10^{-9}	$		\\	
$		$	&	$	|\C{10}|	$	&	$	<6.7\times 10^{-15}	$		\\	
$		$	&	$	|\RC{11}|	$	&	$	<3.8\times 10^{-15}	$		\\	
$		$	&	$	|\IC{11}|	$	&	$	<4.1\times 10^{-15}	$		\\	
\end{tabular}
\caption{Results for spherical coefficients from Hanford, WA preliminary data.}
\end{center}
\end{table}

To calibrate the power spectra,
we take advantage of the strongest tidal line
in the twice-sidereal region,
which is the lunar principle wave M$_2$.
The horizontal gravity gradients from this wave are known
\cite{pm}.
They can be used to calculate the induced phase shift
on the light at the detector port,
given the latitude of the Hanford detector 
and the orientation of the interferometer arms.
This derived value is in close agreement with the result
obtained from the observed modulation of the data
and a simulation of the inteferometer
\cite{am14,finesse}.
Normalizing the spectrum to this phase shift yields
$\De\ph/2\pi = (1.1\pm 1.2)\times 10^{-12}$
for a single traversal at this frequency.
We can use this to extract the values of $\De\ol n/\ol n$
at the various harmonics of $\om_\oplus$ and $\Om_\oplus$
of interest. 

The results of this procedure are shown in Table 4.
The value for each of $\om_\oplus$, $2\om_\oplus$, and $\Om_\oplus$ is 
a $2\si$ confidence limit on a signal above expectation,
while that for each of $3\om_\oplus$ and $4\om_\oplus$ is 
a $2\si$ confidence limit on a signal above noise.
The $16\si$ signal at $2\Om_\oplus$ is obtained from the anomalous peak
discussed above.
In principle,
the phases of the oscillations
and also the various sidebands presented in Table 2
contain interesting information
about Lorentz violation as well.
However,
for the given duration of the run,
the resolution is insufficient
to extract useful information about these sidebands. 

Combining the values in Table 4
with the contributions to $\De\ol n/\ol n$
presented in Tables 1 and 2
yields results for the spherical coefficients for Lorentz violation.
To gain some insight into these results,
we can follow standard procedure in the field
\cite{tables}
and consider the result for each spherical coefficient in turn
under the assumption that all others vanish. 
These results are collected in Table 5.
Additional insight is obtained by working
instead in a cartesian basis.
Results for the $d=4$ cartesian coefficients 
$\kem^{JK}$, $\kop^{JK}$, and $\kt$
are displayed in Table 6.

Overall,
the results in Tables 5 and 6
reveal improvements in laboratory sensitivity 
to all but one of the coefficients for Lorentz violation
associated with operators at $d=4$.
The limits on coefficients controlling parity-even rotation-violating operators 
represent a gain of about four orders of magnitude
over existing laboratory bounds
\cite{tables,ch16,na15,pr15},
while those on parity-odd operators 
are improved by about a factor of four.
In contrast, 
the constraint on $\kt$ in Table 6
is weaker than the best existing two-sided bounds 
from laboratory experiments
\cite{tables,al09,ho09}
and from astrophysics
\cite{tables,sc13}.
Moreover,
a definitive measurement of 
the $d=4$ coefficient $\C{00}$ or, equivalently, $\kt$
cannot be inferred from these results as 
the constraint obtained from the annual frequency $\Om_\oplus$
appears incompatible with the observed signal from the anomalous peak
at $2\Om_\oplus$.
Assuming an appropriate phase at this frequency
yields the results 
$|\C{00}| = (11.1\pm 0.7)\times 10^{-9}$
and $|\kt|= (3.1\pm 0.2)\times 10^{-9}$.
This anomalous signal could conceivably 
be a theoretical artifact of the analysis performed here,
which assumes conventional fermions 
and therefore is insensitive to matter-sector coefficients 
producing distinct effects at the annual and semiannual frequencies
\cite{km13},
but the possibility of an instrumental systematic
means that a compelling resolution of this discrepancy
is unlikely to be attained in the absence of new data. 
The results in Table 5 also represent the first laboratory bounds
obtained on the coefficients $\Cs{jm}$,
albeit at a reduced sensitivity
compared to limits found in studies of the dispersion of light 
from astrophysical sources
\cite{ki15}.

The striking improvement in sensitivity to photon-sector Lorentz violation
revealed in the above analysis suggests that further searches 
using existing gravitational-wave detectors 
would be well worthwhile.
Substantial further gains in sensitivity are likely to be attainable 
by implementing several options.
One is to incorporate results from sites other than Hanford,
including those for LIGO, Virgo, planned ground-based observatories, 
and perhaps eventually space-based missions such as LISA.
A combined analysis would not only increase statistics 
and potentially eliminate systematics 
but would also gain from the differing colatitudes 
and orientations of the instruments.
For example,
a calculation of the contributions from various harmonics 
at the Livingston site 
reveals that the semiannual signal 
is enhanced by a factor of 4.2 due to the geometry of the site, 
which should permit confirmation or refutation of the anomalous peak.
Another potential plus is the improved noise control
now in place for the advanced LIGO instrument,
which could imply a gain in sensitivity to Lorentz violation as well.

\begin{table}
\begin{center}
\setlength{\tabcolsep}{5pt}
\renewcommand{\arraystretch}{1.2}
\begin{tabular}{c|c|c}
	Harmonic		&		Coefficient		&		Result		\\	\hline\hline	
$	\om_\oplus	$	&	$	|\kem^{XZ}|	$	&	$	<2.1\times 10^{-20}	$	\\		
$		$	&	$	|\kem^{YZ}|	$	&	$	<2.1\times 10^{-20}	$	\\	[5pt]	
$	2\om_\oplus	$	&	$	|\kem^{XY}|	$	&	$	<2.7\times 10^{-22}	$	\\		
$		$	&	$	|\kem^{XX}-\kem^{YY}|	$	&	$	<5.5\times 10^{-22}	$	\\	[5pt]	
$	\Om_\oplus	$	&	$	|\kt|	$	&	$	<9.2\times 10^{-10}	$	\\		
$		$	&	$	|\kop^{XY}|	$	&	$	<6.6\times 10^{-15}	$	\\		
$		$	&	$	|\kop^{XZ}|	$	&	$	<5.7\times 10^{-15}	$	\\		
$		$	&	$	|\kop^{YZ}|	$	&	$	<5.2\times 10^{-15}	$	\\
\end{tabular}
\caption{Results for minimal cartesian coefficients from Hanford, WA.}
\end{center}
\end{table}

To summarize,
we have shown in this work
that the interferometers in gravitational-wave observatories
can be used to perform exquisitely sensitive tests of Lorentz invariance
in the photon sector,
thereby extending the role of these instruments 
beyond the more direct searches for Lorentz violation 
in the propagation of gravitational waves
\cite{gravwave}
and other prospective studies of quantum gravity
\cite{other}.
Searches of this type have a reach 
for photon-sector Lorentz violation
that is several orders of magnitude
beyond existing laboratory tests
\cite{tables,ch16,na15,pr15},
and substantial improvements in the results reported here can be envisaged. 
The future is evidently bright for improved studies of Lorentz invariance
in the spirit of the pioneering Michelson-Morley experiment.

\section*{Acknowledgments}

The preliminary data discussed here were obtained during the LIGO S5 run.
We are indebted to the staff and operators of the Hanford LIGO Observatory 
and to the members of the LIGO Scientific Collaboration 
for their dedicated efforts before and during the run. 
In particular,
we thank W.E.\ Butler, C.\ Forrest, T.\ Fricke, 
S.\ Giampanis, F.J.\ Raab, and D.\ Sigg,
who designed, installed, and operated the fsr channel and analyzed the data.
LIGO is operated by Caltech and MIT and is funded 
by the United States National Science Foundation.
The authors were supported in part 
by the United States Department of Energy
under grants {DE}-SC0010120 and DE-FG02-91ER40685,
by the United States National Science Foundation 
under grants PHY-1520570 and PHY-0456239,
and by the Indiana University Center for Spacetime Symmetries.

\end{document}